\documentclass[twocolumn,showpacs,preprintnumbers,amsmath,amssymb]{revtex4}
\usepackage{graphicx}% Include figure files
\usepackage{dcolumn}% Align table columns on decimal point
\usepackage{bm}% bold math
\usepackage{epsfig}
\usepackage{epsfig}
\usepackage{amsmath}
\usepackage{amsfonts}
\usepackage{subfigure}

\newcommand{\ba}{\begin{eqnarray}}
\newcommand{\ea}{\end{eqnarray}}
\newcommand{\be}{\begin{equation}}
\newcommand{\ee}{\end{equation}}
\newcommand{\bea}{\begin{eqnarray}}
\newcommand{\eea}{\end{eqnarray}}

\usepackage{theorem}
\newtheorem{theorem}{Theorem}

\theorembodyfont{\rmfamily}

\theoremstyle{break}

\def\QED{~\rule[-1pt]{5pt}{5pt}\par\medskip}

\def\n{\noindent}

\begin{document}

%\preprint{APS/123-QED}

\title{ Efficient synthesis of quantum gates on indirectly coupled spins }

\author{Haidong Yuan$^1$\footnotemark[1],Daxiu Wei$^2$,Yajuan Zhang$^2$, Steffen Glaser$^3$,Navin Khaneja$^4$ }

\affiliation{ $^1$Department of Applied Mathematics,
The Hong Kong Polytechnic University,Hong Kong\\
%\author{Yajuan Zhang, Daxiu Wei}
$^2$Physics Department and Shanghai Key Laboratory of Magnetic Resonance, East China Normal University, Shanghai 200062, China.\\
%\author{Steffen Glaser}
$^3$Department of Chemistry, Technische Universit$\ddot{a}$t M$\ddot{u}$nchen, Lichtenbergstrasse 4, D-85747 Garching, Germany\\
%\author{Navin Khaneja}
$^4$School of Engineering and Applied Science, Harvard
University,Cambridge, MA 02138}

%\author{Robert Ziert,Navin Khaneja}
%\affiliation{ School of Engineering and Applied Science, Harvard
%University,Cambridge, MA 02138}
\date{\today}% It is always \today, today,

\begin{abstract}
Experiments in coherent nuclear and electron magnetic resonance,
and quantum computing in general correspond to control of quantum
mechanical systems, guiding them from initial to final target
states by unitary transformations. The control inputs (pulse
sequences) that accomplish these unitary transformations should
take as little time as possible so as to minimize the effects of
relaxation and decoherence and to optimize the sensitivity of the
experiments. Here, we derive a time-optimal sequences as fundamental building blocks
for synthesize unitary transformations. Such sequences can be widely implemented on various physical
systems, including the simulation of effective Hamiltonians for topological quantum computing on spin lattices. Experimental demonstrations are provided for a system consisting of three nuclear spins.

\end{abstract}
\maketitle \footnotetext[1]{haidong.yuan@gmail.com}
\section{\label{sec:introduction}Introduction}
The control of quantum systems has important
applications in physics and chemistry. In particular, the ability
to steer the state of a quantum system (or an ensemble of quantum
systems) from a given initial state to a desired target state
forms the basis of spectroscopic techniques such as nuclear
magnetic resonance (NMR) and electron spin resonance (ESR)
spectroscopy \cite{Ernst, electron} and laser coherent control
\cite{optics} and quantum computing \cite{QC1, QC2}.
Experiments in coherent nuclear and electron magnetic resonance,
and optical spectroscopy correspond to control of quantum
mechanical ensembles, guiding them from initial to final target
states by unitary transformations. The control inputs (pulse
sequences) that accomplish these unitary transformations should
take as little time as possible so as to minimize the effects of
relaxation and decoherence and to optimize the sensitivity of the
experiments.
The time-optimal synthesis of unitary operators is now well
understood for coupled two-spin systems \cite{navintoc, Bennett,
Timo2spin, General2spin, yuan, robert}. This problem has also been
recently studied in the context of linear three-spin topologies
\cite{navingeodes, navingeodes_exp, navingeodes2, yuan2, yuan3,Carlini12,Carlini13,Carlini11}. In this article, we use optimal control technique to design pulse sequences to efficiently generate a
class of quantum gates on three-spin systems, and show that such pulse sequences have
 significant savings in the implementation time of trilinear
Hamiltonians and synthesis of couplings between indirectly coupled
qubits over conventional methods. Such pulse sequences also have applications on various other systems.
The article is organized as following: in section \ref{sec:linear}, we first review the previous results on the linearly coupled three-spin system; in section \ref{sec:indirect}, we use optimal control techniques to design the new pulse sequences and show significant savings in the implementation time and the wide applications of such pulse sequences; in section \ref{sec:exp}, we show the experiment implementation of these pulse sequences on NMR; section \ref{sec:conclusion} concludes.
\section{Time optimal control for three linearly coupled spins}\label{sec:linear}
In this section, we give a brief introduction of previous results
on three linearly coupled spins \cite{navingeodes}, to which our
new results is be compared with.

Consider a chain of three spins coupled by scalar couplings
($J_{13} = 0$). Furthermore assume that it is possible to
selectively excite each spin (perform one qubit operations in
context of quantum computing). The goal is to produce a desired
unitary transformation $U \in SU(8)$, from the specified couplings
and single spin operations in shortest possible time. The unitary
propagator $U$, describing the evolution of the system in a
suitable rotating frame is well approximated by

%{\bf (please include spin labels in circles:   1    2    3)}
\begin{figure}[h]
\begin{center}
\includegraphics[scale=.5]{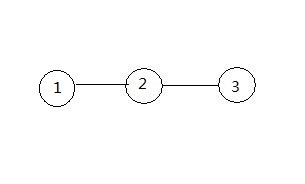}
\end{center}
\caption{ Three linearly coupled spins }
   \label{fig:1}
\end{figure}

\begin{equation} \label{eq:het.three}
\dot{U} = -i (\ H_d + \sum_{j=1}^{6}u_j H_j \ )U , \ \ U(0)=
I \end{equation} where \begin{eqnarray*}H_d &=& 2\pi J_{12} I_{1z} I_{2z} + 2 \pi J_{23}I_{2z}I_{3z}, \\
H_1 &=& 2\pi I_{1x}, \\
\ H_2 &=& 2\pi I_{1y}, \\
\ H_3 &=& 2\pi I_{2x}, \\
H_4 &=& 2\pi I_{2y}, \\
H_5 &=& 2 \pi I_{3x}, \\
H_6 &=& 2 \pi I_{3y}.
\end{eqnarray*}
We use the notation $I_{\ell \nu}= \bigotimes_{j} I_{a_{j}}$,
where $a_{j}=\nu$ for $j=\ell$ and $a_{j}=0$ otherwise  (see
\cite{Ernst}). The matrices $I_{x}:= \frac{1}{2}\left(\begin{smallmatrix}
0 & 1 \\
1 & 0
\end{smallmatrix}
\right)$, $ I_{y}:= \frac{1}{2}\left(
\begin{smallmatrix}
0 & -i \\
i & 0
\end{smallmatrix}
\right) $, and $ I_{z}:= \frac{1}{2}\left(
\begin{smallmatrix}
1 & 0 \\
0 & -1
\end{smallmatrix}
\right) $, are the Pauli spin matrices and $ I_{0}:= \left(
\begin{smallmatrix}
1 & 0 \\
0 & 1
\end{smallmatrix}
\right) $ is the $2\times 2$-dimensional identity matrix. The
symbol $J_{12}$ and $J_{23}$ represents the strength of scalar
couplings between spins $(1,2)$ and $(2,3)$ respectively, here we
will treat the important case of this problem when the couplings
are both equal ($J_{12} = J_{23} = J $). We will be most
interested in a unitary propagator of the form $$U = \exp(-i
\theta \ I_{1z}I_{2z}I_{3z}).$$ These propagators are hard to
produce as they involve trilinear terms in the effective
Hamiltonian. We will refer to such propagators as {\it trilinear
propagators}.

We assume that we can selectively rotate each spin at a rate much
faster than the evolution of the couplings, i.e., the single spin
operations can be done in negligible time.

\begin{theorem}\label{th:main.0}\cite{navingeodes}{\rm \ Given the spin system in (\ref{eq:het.three}),
with $J_{12} = J_{23} = J$ and $J_{13}=0$, the minimum time
$t^{\ast}(U_F)$ required to produce a propagator of the form $U_F
= \exp(-i \theta I_{1z}I_{2z}I_{3z}),\ \ \theta \in [0, 4 \pi] $
is given by
\begin{equation}
\label{eq:tstar}
t^{\ast}(U_F) = \frac{\sqrt{2\pi \theta - (\theta/2)^{2}}}{2 \pi J} = \frac{\sqrt{\kappa(4 - \kappa)}}{2J},
\end{equation}
where $\kappa = \frac{\theta}{2 \pi}$.} The pulse sequence that
produces the propagator $U_F$ is as follows.
\bea \aligned
 U_F &= \exp(-i \frac{\pi}{2}I_{2y}) \ \exp(-i [\pi +
\frac{\beta}{2}] I_{2x}) \\
& \exp(T [-i 2 \pi J (I_{1z}I_{2z} + I_{2z}I_{3z}) + i
\frac{\beta}{T}I_{2x}]) \exp( i \frac {\pi}{2}I_{2y})
\endaligned
\eea Where $\beta=2\pi-\theta/2$ and $T=\frac{\sqrt{\kappa(4 -
\kappa)}}{2J}$.
\end{theorem}

\section{Quantum gates between indirectly coupled spins}\label{sec:indirect}
\subsection{Time optimal sequences}
In this section we derive a time optimal sequence for a control problem arises from simulation of trilinear couplings of indirectly coupled spins. First we consider to generate the trilinear term $e^{\theta S_1}$, where
$S_1=-4i [I_{1x}I_{2z}I_{3y}+I_{1y}I_{2z}I_{3x}]$.

Let
\begin{eqnarray}\label{eq:S}
\aligned
%S_1&=-4i[I_{1x}I_{2z}I_{3y}+I_{1y}I_{2z}I_{3x}],\\
S_2&=-2i[I_{1x}I_{2x}+I_{2x}I_{3x}], \\
S_3&=-2i[I_{1y}I_{2y}+I_{2y}I_{3y}],\\
\endaligned
\end{eqnarray}
 here $S_2, S_3$ are locally equivalent to the coupling Hamiltonian $H_d$(up to a re-scaling of time, from now on we assume the time unit is $\frac{1}{\pi J}$),
 $$S_2= e^{-\frac{\pi}{2} i(I_{1y}+I_{2y}+I_{3y}}(-iH_d)e^{\frac{\pi}{2} i(I_{1y}+I_{2y}+I_{3y}},$$
  $$S_3= e^{\frac{\pi}{2} i(I_{1x}+I_{2x}+I_{3x}}(-iH_d)e^{-\frac{\pi}{2} i(I_{1x}+I_{2x}+I_{3x}}.$$

Notice that $S_1, S_2, S_3$ form an $so(3)$ algebra, i.e.
$$[S_1,S_2]=S_3, $$
$$[S_2, S_3]=S_1,$$
$$[S_3,S_1]=S_2,$$
thus we can map $S_1, S_2, S_3$ to $\Omega_x,\Omega_y, \Omega_z$, where
$$\Omega_x=\left(\begin{array}{ccc}
      0 & 0 & 0 \\
      0 & 0 & -1 \\
      0 & 1 & 0
\end{array}\right).$$
 $$\Omega_y=\left(\begin{array}{ccc}
      0 & 0 & 1 \\
      0 & 0 & 0 \\
      -1 & 0 & 0
\end{array}\right),$$
$$\Omega_z=\left(\begin{array}{ccc}
      0 & -1 & 0 \\
      1 & 0 & 0 \\
      0 & 0 & 0
\end{array}\right).$$
Then we reformulate the problem as following:
\begin{equation}
\label{eq.omega}
\frac{d}{dt}\Omega=A\Omega,
\end{equation}
$$A=u(t)\Omega_z+v(t)\Omega_y,$$
 here $u(t),v(t)$ can be $\{\pm 1,0\}$ as we can change the sign of the Hamiltonian by local controls, and it is $0$ when it is not switched on. Also at each instant of time, only one Hamiltonian can be switched on, so $\forall t, |u(t)|+|v(t)|=1$.

Now the question becomes how to generate $\Omega(T)=e^{\alpha\Omega_x}$ in minimum time under the dynamics governed by Eq.(\ref{eq.omega}) with the initial condition $\Omega(0)=I$.
This is equivalent to find the optimal sequences $$\exp(\pm \Omega_yt_1)\exp(\pm \Omega_zt_2)\exp(\pm \Omega_yt_3)\exp(\pm \Omega_zt_4)\cdots$$ to generate $\exp(\alpha \Omega_x)$,
such that $T=\sum_i t_i$ is minimized.
The conventional method is to use the Baker-Campbell-Hausdorff(BCH) formula
$$e^{\alpha\Omega_x}=e^{\frac{\pi}{2}\Omega_y}e^{\alpha\Omega_z}e^{-\frac{\pi}{2}\Omega_y},$$ which corresponds to set $u(t)=0, v(t)=1$ when $t\in[0,\frac{\pi}{2}]$, $u(t)=1, v(t)=0$ when $t\in[\frac{\pi}{2},\frac{\pi}{2}+\alpha]$ and $u(t)=0, v(t)=-1$ when $t\in[\frac{\pi}{2}+\alpha,\pi+\alpha]$, the total duration is $\pi+\alpha$ units of time. While this is optimal when $\alpha$ is larger than $\frac{\pi}{2}$, for small $\alpha$, it is far from optimal since there is always an offset $\pi$ for the total time.

The minimum time can actually be achieved by adding one switch.
% reduce the time cost significantly, especially when $\alpha$ is small, thus improve the fidelity of these gates by reducing the time exposed to the noises.
%The sequence consists of three switches,
For $\alpha\in[0,\frac{\pi}{2}]$, the time optimal sequence takes the form
\begin{equation}\label{eq:sequence}
e^{\alpha \Omega_x}=e^{t_2\Omega_z}e^{-\delta t \Omega_y}e^{-\delta t\Omega_z}e^{t_1\Omega_y},
\end{equation}
while
\begin{eqnarray}
\aligned
t_2=t_1&=arccos\frac{1}{\sin\frac{\alpha}{2}+\cos\frac{\alpha}{2}},\\
\delta t&=arccos(\cos\frac{\alpha}{2}-\sin\frac {\alpha}{2}).
\endaligned
\end{eqnarray}

The total time to generate $e^{\alpha \Omega_x}$ is
\begin{equation}\label{Eq:time}
f(\alpha)=2[arccos\frac{1}{\sin\frac{\alpha}{2}+\cos\frac{\alpha}{2}}+arccos(\cos\frac{\alpha}{2}-\sin\frac {\alpha}{2})].
\end{equation}
Symmetrically when $\alpha\in [-\frac{\pi}{2},0]$, $$f(\alpha)=f(-\alpha).$$ So for $\alpha \in [-\frac{\pi}{2},\frac{\pi}{2}]$,
\begin{equation}
\nonumber
f(\alpha)=2[arccos\frac{1}{\sin\frac{|\alpha|}{2}+\cos\frac{|\alpha|}{2}}+arccos(\cos\frac{|\alpha|}{2}-\sin\frac{|\alpha|}{2})].
\end{equation}

\begin{figure}[h]
\begin{center}
\includegraphics[scale=0.4]{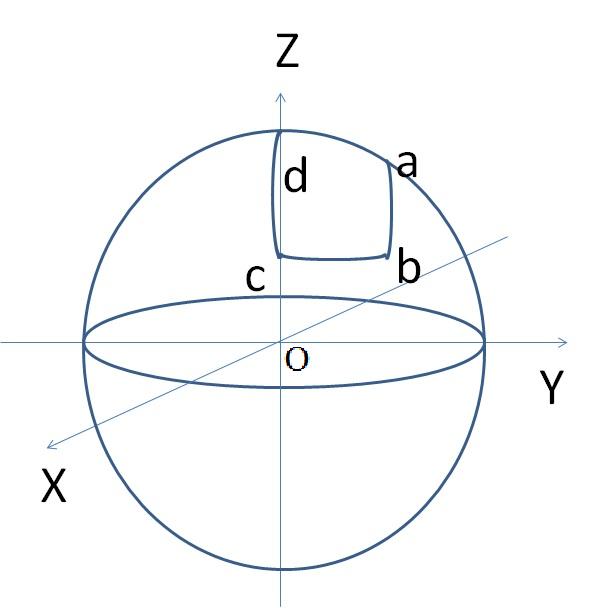}
\end{center}
\caption{ Rotation corresponds to $e^{\alpha\Omega_x}$ that moves $a=(0, \sin\alpha, \cos\alpha)$ to $d=(0,0,1)$ via the point $b=(\cos\alpha \sin t_1, \sin\alpha, \cos\alpha \cos t_1)$ and $c=(\sqrt{\cos^2\alpha \sin^2 t_1+\sin^2\alpha},0,\cos\alpha \cos t_1)$.}
   \label{fig:rotation}
\end{figure}
 The detailed derivation of this sequence can be found in the appendix, here we just give an intuitive picture of how the sequence works. Geometrically $\{e^{\alpha\Omega_x}\}$ corresponds to the a rotation on a sphere which moves the point $(0, \sin\alpha, \cos\alpha)$ to (0,0,1) while keeps the x-axis fixed. The sequence in Eq.(\ref{eq:sequence}) corresponds to four steps that achieve such a rotation. As shown in Fig.\ref{fig:rotation}, it is first rotated around the y-axis for $t_1$ time, which rotates the point $a=(0, \sin\alpha, \cos\alpha)$ to $b=(\cos\alpha \sin t_1, \sin\alpha, \cos\alpha \cos t_1)$; then it's rotated around (-z)-axis for $\delta t$ time, which rotates point $b$ to the point $c=(\sqrt{\cos^2\alpha \sin^2 t_1+\sin^2\alpha},0,\cos\alpha \cos t_1)$ which is on the XZ-plane; after that it is rotated around (-y)-axis for $\delta t$ time to the point $d=(0,0,1)$; since x-axis is orthogonal to $\overrightarrow{oa}$ and rotations does not change the angles, so after these rotations, the x-axis has moved to somewhere which is orthogonal to $\overrightarrow{od}$, i.e., somewhere on the XY-plane, so one needs to make another rotation around the z-axis for $t_2$ time, that moves the x-axis back to the original position.

 It needs to be noted that when writing the manuscript, it was brought to the authors attention that Dirk Mittenhuber had a similar construction of the time optimal sequences\cite{Mittenhuber}.
\subsection{Quantum gates on indirectly coupled spins}
\begin{figure}[h]
\begin{center}
         \includegraphics[scale=.35]{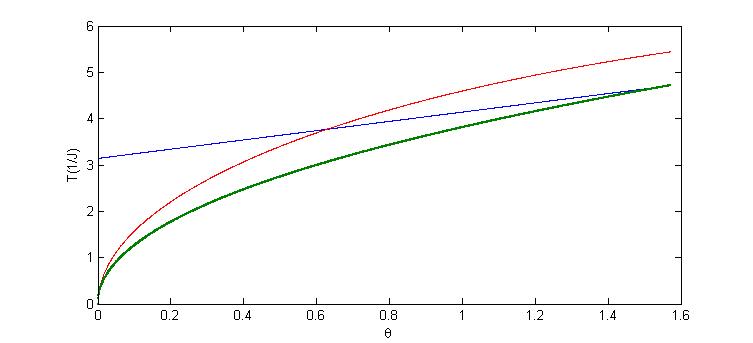}
\end{center}
\caption{ The green line represents the total time generating $e^{-i4\theta [I_{1x}I_{2z}I_{3y}+I_{1y}I_{2z}I_{3x}]}$ using the optimal sequences, the red line is the total time required by the strategy discussed in section \ref{sec:linear},which is just twice the time $t^*(U_F)$ as in Eq.~\ref{eq:tstar}, and the blue line is the total time required by using BCH formula.}
   \label{fig:time}
\end{figure}

 The sequence can be directly used to simulate the trilinear terms $e^{\theta S_1}$ on the linearly coupled three-spin system, which is
\begin{equation}\label{eq:spinsequence}
e^{\alpha S_1}=e^{t_2S_3}e^{-\delta t S_2}e^{-\delta tS_3}e^{t_1S_2},
\end{equation}
where $S_1,S_2,S_3$ are as defined in last section. The minimum time needed to generate $e^{\theta S_1},\theta\in [-\frac{\pi}{2},\frac{\pi}{2}],$ is
\begin{equation}
f(\theta)=2[arccos\frac{1}{\sin\frac{|\theta|}{2}+\cos\frac{|\theta|}{2}}+arccos(\cos\frac{|\theta|}{2}-\sin\frac {|\theta|}{2})].
\end{equation}
 It is not only shorter than the conventional method using BCH formula, but is also shorter than the strategy described in section.\ref{sec:linear}. In Fig. ~\ref{fig:time}, we plotted the total time needed for different strategies.
% using the optimal sequence together with the total time using the strategy in section.\ref{sec:linear}---which is just twice the time $t^*(U_F)$ as in Eq.~\ref{eq:tstar}.
%\section{Application}\label{sec:application}
%\subsection{Induced gates between indirected coupled spins}

\begin{figure}[h]
\begin{center}
\includegraphics[scale=.5]{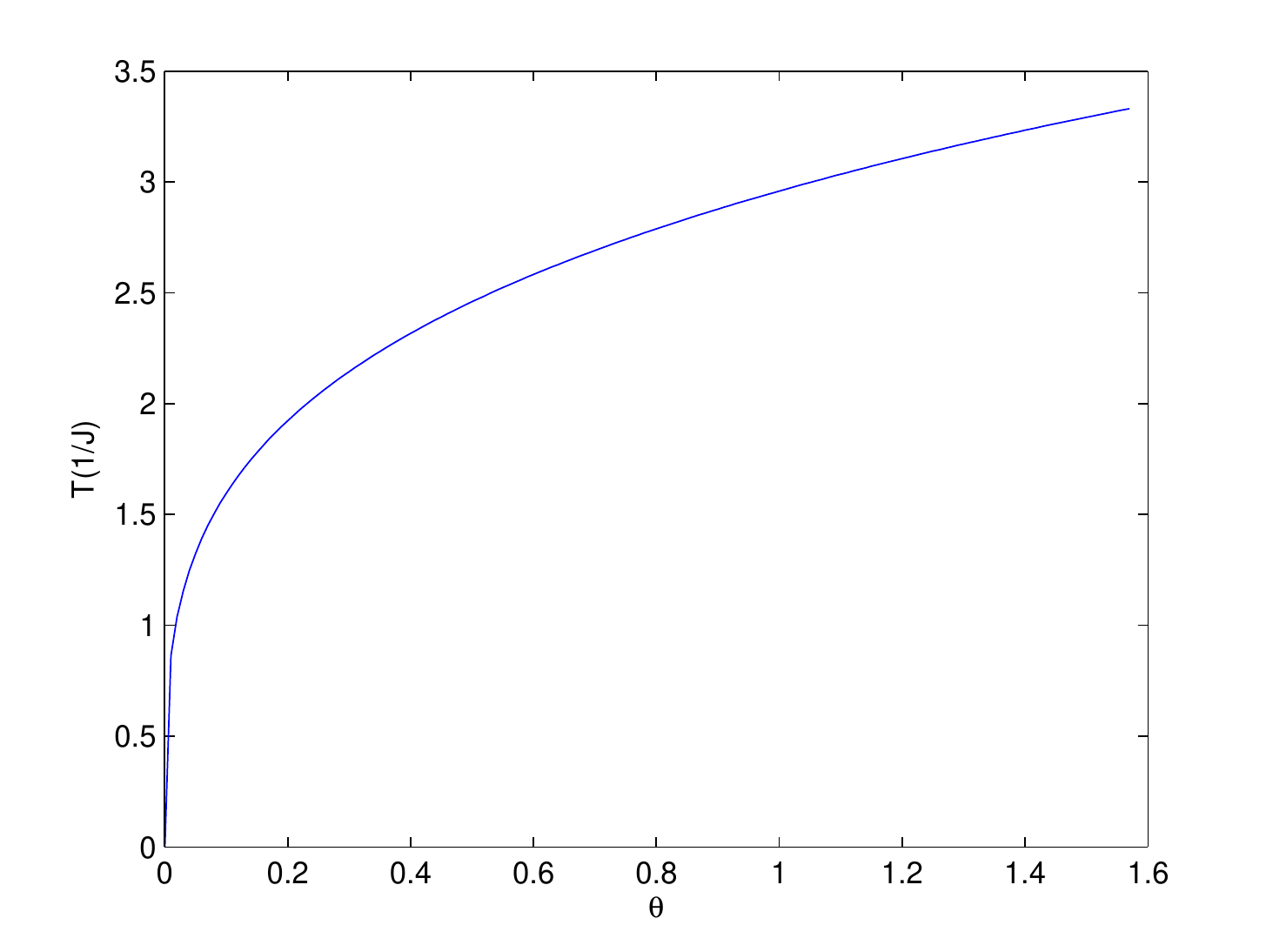}
\end{center}
\caption{ Time needed to generate $e^{-2i\theta[I_{1x}I_{3x}+I_{1y}I_{3y}]}$ }
   \label{fig:g}
\end{figure}

We can also construct various other gates between indirected coupled spin 1 and spin 3 in the linearly coupled three-spin system by concatenating the optimal sequence. For example, we can  build gates $e^{-2i\theta[I_{1x}I_{3x}+I_{1y}I_{3y}]}, \theta\in (0,\frac{\pi}{2})$ on spin 1 and 3. Since
\begin{eqnarray}
\aligned
S_1&=-4i [I_{1x}I_{2z}I_{3y}+I_{1y}I_{2z}I_{3x}],\\
S_4&=-2i[I_{1z}I_{2z}+I_{2z}I_{3z}],\\
S_5&=-2i[I_{1x}I_{3x}-I_{1y}I_{3y}].\\
\endaligned
\end{eqnarray}
 form an $so(3)$ algebra, we can use the optimal sequence to get
\begin{eqnarray}
%\aligned
e^{\theta S_5}=e^{-t_2S_1}e^{-\delta t S_4}e^{\delta t S_1}e^{t_1 S_4},
%\endaligned
\end{eqnarray}
where \begin{eqnarray}
\aligned
t_1=t_2&=arccos\frac{1}{\sin\frac{\theta}{2}+\cos\frac{\theta}{2}},\\
\delta t&=arccos(\cos\frac{\theta}{2}-\sin\frac {\theta}{2}).
\endaligned
\end{eqnarray}
Here $e^{-t_2S_1}$ and $e^{\delta t S_1}$ can again be decomposed into four pulses, we can thus generate $e^{\theta S_5}=e^{-2i\theta[I_{1x}I_{3x}-I_{1y}I_{3y}]}$ with the total time $g(\theta)=t_1+\delta t+f(t_1)+f(\delta t)$(plotted in Fig.\ref{fig:g}), where $f$ is the function given in Eq.(\ref{Eq:time}). This is also the total time to generate $e^{-2i\theta[I_{1x}I_{3x}+I_{1y}I_{3y}]}$ as $$e^{-2i\theta[I_{1x}I_{3x}+I_{1y}I_{3y}]}=e^{i\pi I_{3x}}e^{\theta S_5}e^{-i\pi I_{3x}}.$$

Similarly we can efficiently build the gates
\begin{eqnarray}
%\nonumber
%\aligned
%e^{-i\theta[I_{1x}I_{3x}+I_{1y}I_{3y}]},\\
\nonumber
e^{-2i\theta[I_{1x}I_{3x}+I_{1z}I_{3z}]}\\
\nonumber
\end{eqnarray}
and
\begin{eqnarray}
e^{-2i\theta[I_{1y}I_{3y}+I_{1z}I_{3z}]}.\\
\nonumber
\end{eqnarray}
Combining them, we can efficiently simulate the Heisenberg coupling $e^{-2i\theta[I_{1x}I_{3x}+I_{1y}I_{3y}+I_{1z}I_{3z}]}$ between indirectly coupled spins.

%We plot the function $g(\theta)$ in Fig.\ref{fig:g}. %is the time needed to generate $e^{-2i\theta[I_{1x}I_{3x}-I_{1y}I_{3y}]}$.

\subsection{Simulation of multi-body interactions}
The sequence can also be used to simulate topological quantum computing on a spin lattice. Topological quantum computing on a spin lattice requires multi-body interaction Hamiltonians, for example, the toric code model on a square lattice requires four-body interaction, and the honeycomb lattice model requires six-body interaction. Most physical Hamiltonians only contain two-body interaction. The pulse sequence can be used to simulate multi-body Hamiltonian on physical systems, which promises faster execution and higher fidelity.

Let's consider the toric code on square lattice, the desired Hamiltonian is given by
$$H_T=-J_e\sum_sA_s-J_m\sum_pB_p$$
where $s$ runs over vertices (stars) of the lattice and $p$ runs over the plaquettes.
\begin{figure}[h]
\begin{center}
\includegraphics[scale=1]{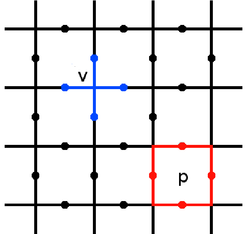}
\end{center}
\caption{ The spins live on the edges of the square lattice. The
spins adjacent to a star operator As and a plaquette operator Bp are shown }
   \label{fig:1}
\end{figure}

The star operator acts on the four spins surrounding a vertex $s$, $$A_s=\prod_{j\in star(s)}\sigma_j^x$$
 while the plaquette operator acts on the four spins surrounding a plaquette,
$$B_p=\prod_{j\in \partial p}\sigma_j^z$$
where $\sigma_{x}:= \left(\begin{smallmatrix}
0 & 1 \\
1 & 0
\end{smallmatrix}
\right)$, $ \sigma_{y}:= \left(
\begin{smallmatrix}
0 & -i \\
i & 0
\end{smallmatrix}
\right) $, and $ \sigma_{z}:= \left(
\begin{smallmatrix}
1 & 0 \\
0 & -1
\end{smallmatrix}
\right) $.
% and we denote $ I_{0}:= \left(
%\begin{smallmatrix}
%1 & 0 \\
%0 & 1
%\end{smallmatrix}
%\right) $ as the $2\times 2$-dimensional identity matrix.
%
Assume we have two-body Ising-coupling between nearest neighbor and selective fast single spin operations, i.e., we have $\sigma_z\otimes\sigma_z$ couplings between two neighboring spins. We show how to use this to simulate the toric code model. We just show how to generate $\exp(-i\theta B_p)$, i.e., simulate one $B_p$ term, as all the terms commute with each other, they can be simulated in a similar way parallelly.

Let's first index the four spins $B_p$ act on with number $1,2,3,4$, and assume they have couplings to nearest neighbor: $\sigma_{1z}\sigma_{2z}$, $\sigma_{2z}\sigma_{3z}$, $\sigma_{3z}\sigma_{4z}$, $\sigma_{4z}\sigma_{1z}$, where $\sigma_{1z}\sigma_{2z}$ denote an operator acting $\sigma_z$ on spin $1$ and $2$ and identity on other spins. To simulate $\exp(-i\theta B_p)$, we can implement the time optimal sequence as following:
\begin{widetext}
\begin{equation}
\exp(-i\theta B_p)=\exp(-it_1\sigma_{1z}\sigma_{2z}\sigma_{3y})\exp(it_2\sigma_{3x}\sigma_{4z})\exp(it_2 \sigma_{1z}\sigma_{2z}\sigma_{3y})\exp(-it_1\sigma_{3x}\sigma_{4z}),
\end{equation}
\end{widetext}
where
\begin{eqnarray}
\aligned
t_1&=arccos\frac{1}{\sin\frac{\theta}{2}+\cos\frac{\theta}{2}},\\
t_2&=arccos(\cos\frac{\theta}{2}-\sin\frac {\theta}{2}).
\endaligned
\end{eqnarray}
Here $\exp(-it_1 \sigma_{1z}\sigma_{2z}\sigma_{3y})$ can be generated by the sequence
\begin{widetext}
\begin{equation}
\exp(-it_1 \sigma_{1z}\sigma_{2z}\sigma_{3y})=\exp(-it_1'\sigma_{1z}\sigma_{2y})\exp(it_2'\sigma_{2x}\sigma_{3y})\exp(it_2' \sigma_{1z}\sigma_{2y})\exp(-it_1'\sigma_{2x}\sigma_{3y}),
\end{equation}
\end{widetext}
where terms containing $\sigma_x$ and $\sigma_y$ are obtained from single spin operations on Ising coupling and
\begin{eqnarray}
\aligned
t_1'&=arccos\frac{1}{\sin\frac{t_1}{2}+\cos\frac{t_1}{2}},\\
t_2'&=arccos(\cos\frac{t_1}{2}-\sin\frac {t_1}{2}).
\endaligned
\end{eqnarray}
And the term $\exp(it_2 \sigma_{1z}\sigma_{2z}\sigma_{3y})$ can be similarly generated. Thus by repeatedly using the four-pulses sequence, one can efficiently simulate $B_p$. Please note that the concatenation of the optimal sequence may not keep its optimality, nevertheless it is much more efficient than conventional method using BCH formula.

\section{Experiment}\label{sec:exp}
\begin{figure}[h]
\begin{center}
\includegraphics[scale=.5]{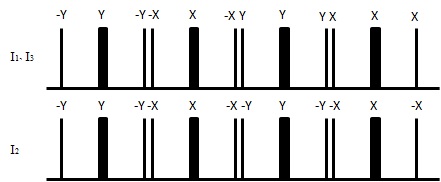}
\end{center}
\caption{
The pulse sequence diagram for the unitary transformation $e^{2\pi i(I_{1x}I_{2z} I_{3y}+I_{1y} I_{2z} I_{3x})}$, where thin vertical lines denote 90$^\circ$ pulses and 180$^\circ$ pulses (the wide vertical lines) are inserted for refocusing of frequency offset effects. The durations $t_1$,$t_2$ and $\delta t$ can be calculated according to the above theory. That is $t_1=t_2=arccos\frac{1}{\sin\frac{\theta}{2}+\cos\frac{\theta}{2}},
\delta t=arccos(\cos\frac{\theta}{2}-\sin\frac {\theta}{2})$. }
   \label{fig:pulses}
\end{figure}

\begin{figure}[ht]
\label{fig:exp}
\begin{center}
\subfigure[Initial state  $I_{1x}$]{\includegraphics[scale=.1]{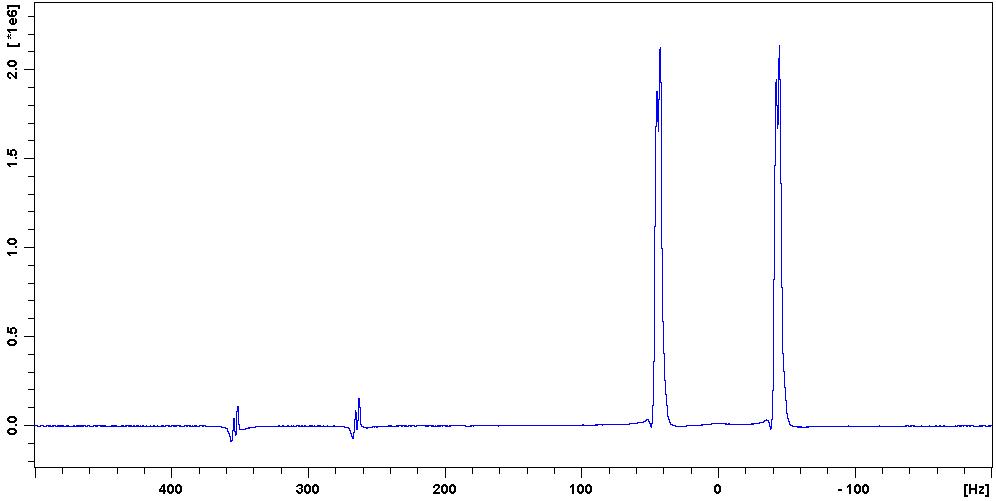}}
\subfigure[Final state $4I_{1z}I_{2z}I_{3x}$]{\includegraphics[scale=.1]{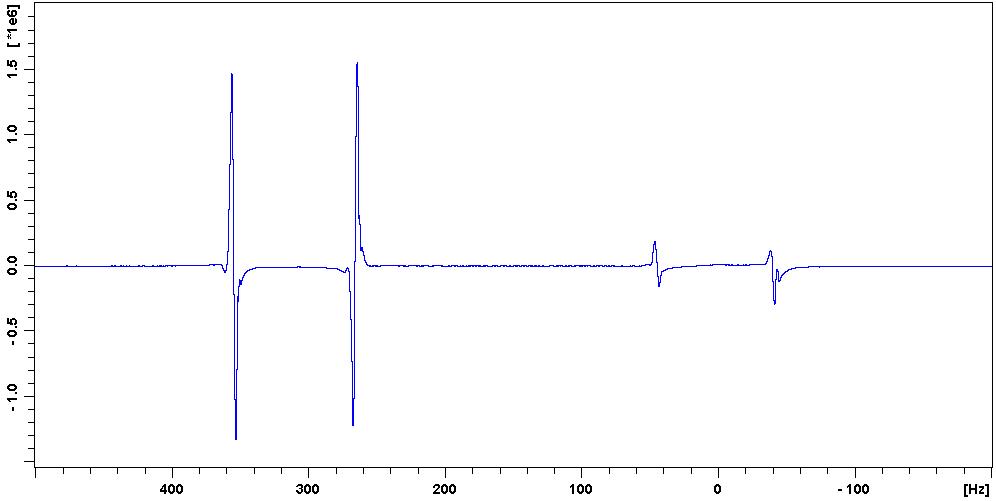}}\\
\subfigure[Initial state  $I_{1y}$]{\includegraphics[scale=.1]{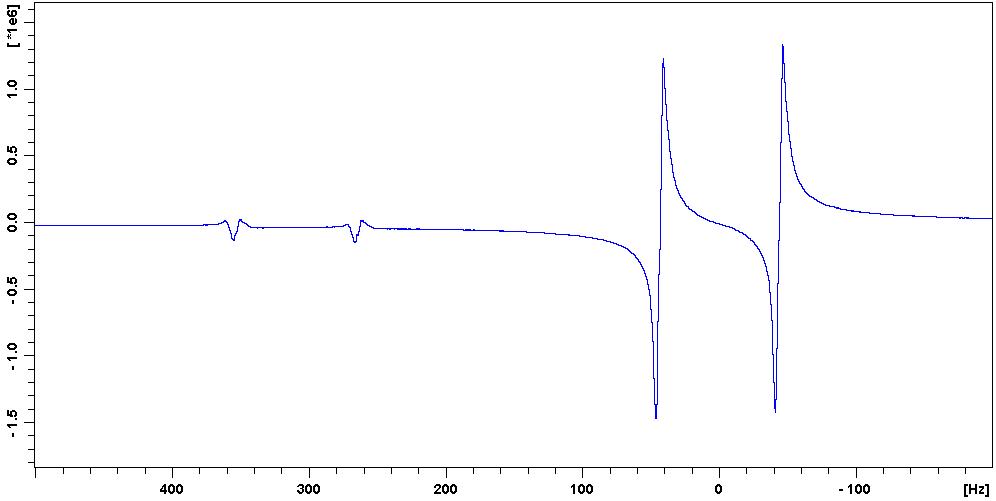}}
\subfigure[Final state $-4I_{1z}I_{2z}I_{3y}$]{\includegraphics[scale=.1]{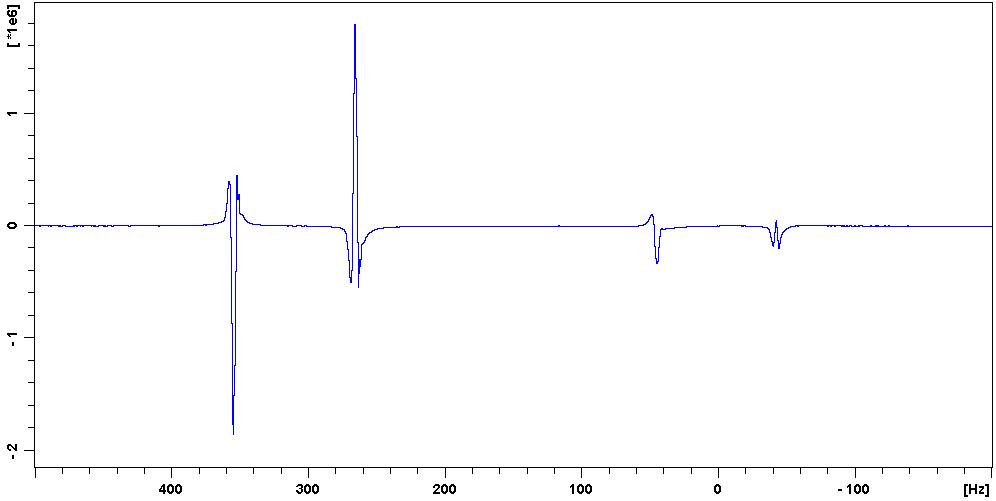}}\\
\subfigure[Initial state  $I_{3x}$]{\includegraphics[scale=.1]{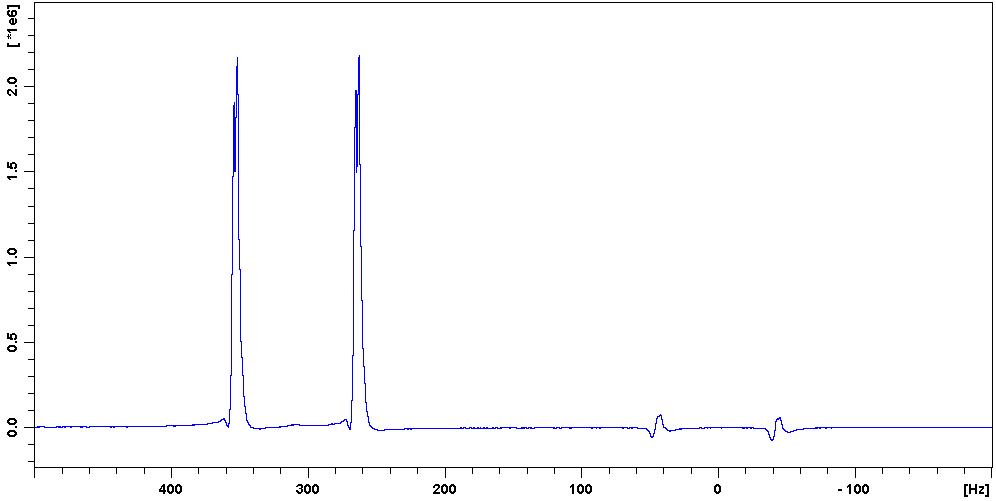}}
\subfigure[Final state $4I_{1x}I_{2z}I_{3z}$]{\includegraphics[scale=.1]{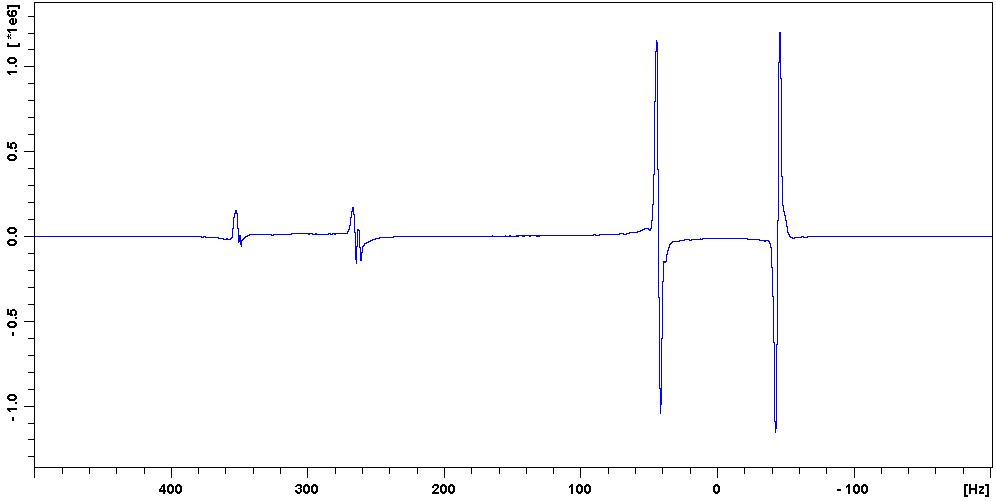}}\\
\subfigure[Initial state  $I_{3y}$]{\includegraphics[scale=.1]{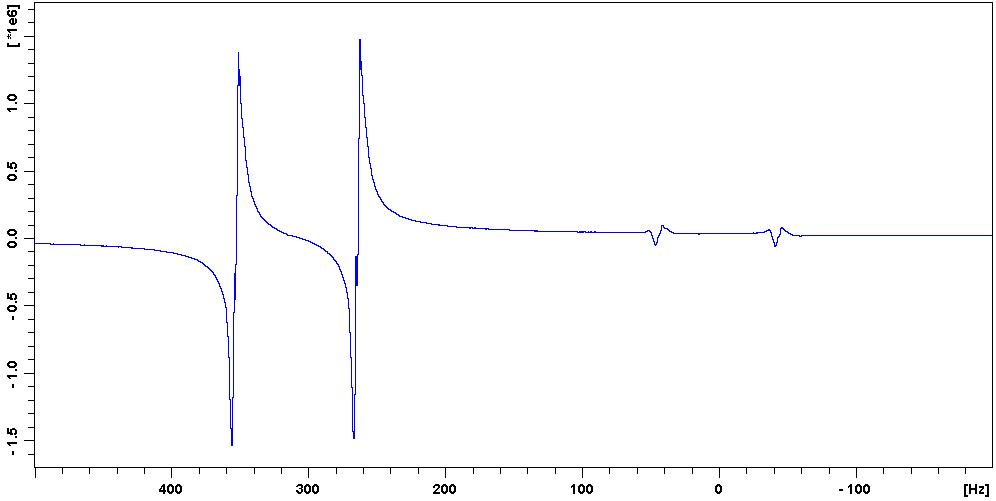}}
\subfigure[Final state $-4I_{1y}I_{2z}I_{3z}$]{\includegraphics[scale=.1]{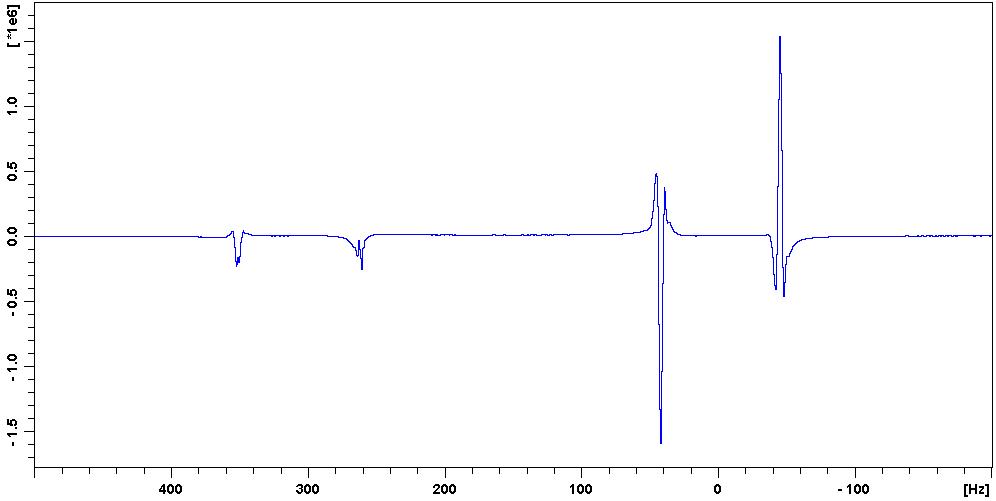}}\\
\subfigure[Initial state  $I_{1z}$]{\includegraphics[scale=0.6]{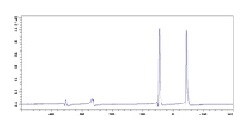}}
\subfigure[Final state $-I_{3z}$]{\includegraphics[scale=.1]{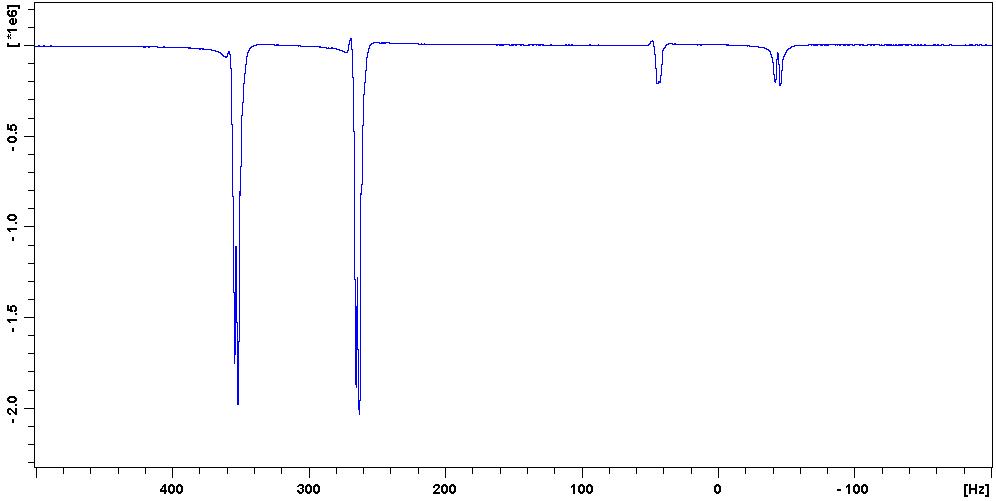}}\\
\subfigure[Initial state  $I_{3z}$]{\includegraphics[scale=.1]{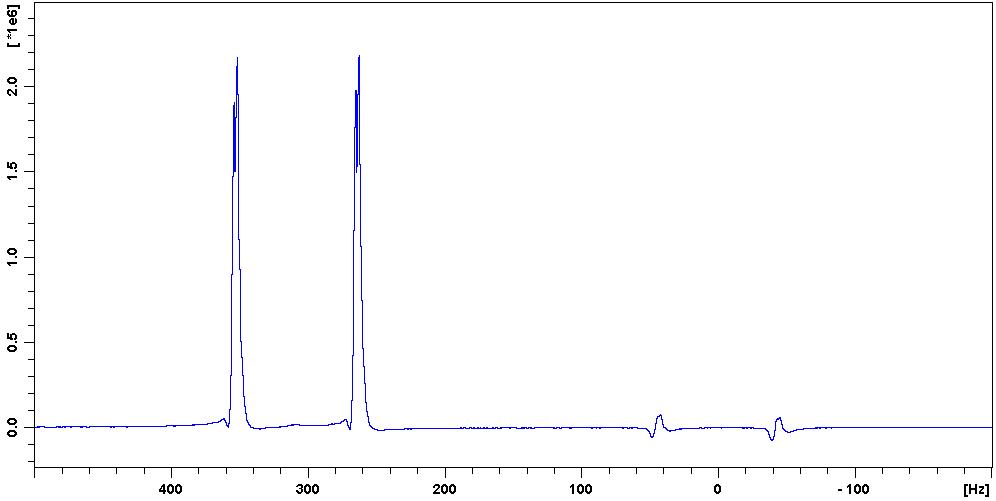}}
\subfigure[Final state $-I_{1z}$]{\includegraphics[scale=.1]{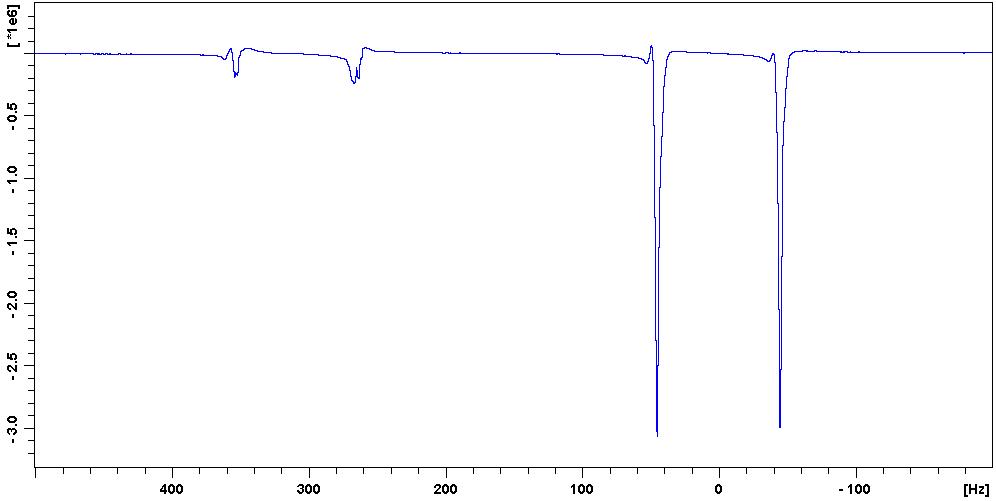}}
%
%
%
%
%\begin{subfigure} [h]{0.1pt}
%\centering
%\includegraphics[scale=.2]{I1x.jpg}
%\caption{Initial state $I_{1x}$}
%\end{subfigure}
%~
%\begin{subfigure} [h]{0.1pt}
%\centering
%\includegraphics[scale=.2]{I1zI2zI3x.jpg}
%\caption{Final state $4I_{1z}I_{2z}I_{3x}$}
%\end{subfigure}
\end{center}
\caption{NMR spectra of different initial and final states, where left column are spectra of different initial states and right column are spectra of corresponding output states.}
\end{figure}

We experimentally demonstrated the implementation of the unitary transformation $e^{2\pi i(I_{1x} I_{2z} I_{3y}+I_{1y} I_{2z} I_{3x})}$ on a BRUKER AVANCE 500M NMR spectrometer. The sample is the amino moiety of 15N acetamide $(NH_2COCH_3)$. Two protons in the spin system $-NH_2$ present the spins 1 and 3. The chemical shift difference between the two protons is 306 Hz. Nuclear 15N denotes spin 2. The J-coupling constants among the three spins are $J_{12}=J_{23}=88Hz$, $J_{13}=2.6Hz$. The pulse sequence used in the experiment is shown in Fig.\ref{fig:pulses}.
In this experiment we take $\theta=2\pi$, $t_1=t_2=2.84ms$, $\delta t=5.68ms$. The whole duration of this pulse is 17ms. We choose six different initial states to observe the final states after the pulses are applied. For the six different initial states $I_{1x}$, $I_{1y}$, $I_{3x}$, $I_{3y}$, $I_{1z}$, $I_{3z}$, the corresponding final states should be $I_{1z}I_{2z}I_{3x}$, $-I_{1z}I_{2z}I_{3y}$, $I_{1x}I_{2z}I_{3z}$, $-I_{1y}I_{2z}I_{3z}$, $-I_{3z}$ and $-I_{1z}$ , respectively.  Fig.\ref{fig:exp} shows the experimental spectra(for the state $I_{iz}$ (i=1, 3), a 90 degree reading pulse along y axis has been applied before acquisition) and Fig.\ref{fig:sim} in the appendix shows the simulations using the theory. The consistency between theory and experimental spectra indicates that the pulse sequence works accurately and the unitary transformation has been experimentally implemented.
\section{Conclusion}\label{sec:conclusion}
We derived a time optimal sequences to generate quantum gates, which can be widely used on various physical systems, for example, it can be used to generate various quantum gates on indirectly coupled spins and can also be used to simulate topological quantum computing on spin lattice.
\begin{acknowledgments}
D. Wei acknowledges the financial support from the National Natural Science Foundation of China(Grant No 11005039).
\end{acknowledgments}
\appendix
\section{Time optimal pulse sequences}
In this Appendix, we derive the solution of the optimal control problem for
\begin{equation}
\frac{d}{dt}\Omega=A\Omega
\end{equation}
$$A=u(t)\Omega_z+v(t)\Omega_y$$
$u(t),v(t)\in \{1, -1, 0\}$ and $\forall t, |u(t)|+|v(t)|=1$. $\Omega(0)=I$, we want to generate $\Omega(T)=e^{\alpha\Omega_z}$ in minimum time. For the rest of the derivation, we will omit $t$ for ease of notation.

We first apply the maximum principle\cite{Pontryagin}. Since the system
has a fat distribution, there is no singular
extremal\cite{Montgomery}. The control hamiltonian then can be written as %(it's the
%hamiltonian in maximal principle, not the physics hamiltonian)
$$H=-1+tr(\lambda ^T A\Omega),$$
where $\lambda$ is an auxiliary 3-dim vector variable and
$\dot{\lambda}=A\lambda$, the optimal control $u,v$ should maximize the control Hamiltonian, i.e., $u,v=argmax\{ H\}$.

\begin{eqnarray}
\aligned
 H&=-1+tr(\lambda ^T A\Omega)\\
  &=-1+tr(A\Omega\lambda ^T)\\
  &=-1+tr(A\frac{\Omega\lambda ^T-\lambda \Omega^T}{2}+A\frac{\Omega\lambda ^T+\lambda \Omega^T}{2}),\\
  &=-1+tr(A\frac{\Omega\lambda ^T-\lambda \Omega^T}{2}),
\endaligned
\end{eqnarray}
where the last step holds because $A$ is skew symmetric and $S=\frac{\Omega\lambda ^T+\lambda \Omega^T}{2}$ is symmetric, as $$tr(AS)=tr[(AS)^T]=tr(-SA)=-tr(AS),$$ thus $tr(AS)=0$. Now let
$$M=\frac{\Omega\lambda ^T-\lambda \Omega^T}{2}=m_x\Omega_x+m_y\Omega_y+m_z\Omega_z,$$
then
\begin{eqnarray}
\aligned
 H=&-1+tr(um_z\Omega_z^2)+tr(vm_y\Omega_y^2)\\
 =&-1-2um_z-2vm_y.
\endaligned
\end{eqnarray}
%$$\dot{\lambda}=A\lambda$$
From the definition of $M$, we get
$$\dot{M}=[A,M],$$
expanding $M$ and $A$, it gives
$$\dot{m_x}\Omega_x+\dot{m_y}\Omega_y+\dot{m_z}\Omega_z=[u\Omega_z+v\Omega_y,m_x\Omega_x+m_y\Omega_y+m_z\Omega_z],$$
we thus get
\begin{eqnarray}
\label{eq:M}
\aligned
\dot{m_x}&=-um_y+vm_z,\\
\dot{m_y}&=um_x,   \\
\dot{m_z}&=-vm_x,\\
\endaligned
\end{eqnarray}
%\end{displaymath}
as $u,v=argmax\{H\}$, we get $$u=-sgn(m_z), v=0 \textrm{  if  } |m_z|>|m_y|,$$
$$u=0, v=-sgn(m_y) \textrm{ if  } |m_z|<|m_y|,$$
if $|m_z|=|m_y|$, then $u,v$ can be either.

\begin{figure}[h]
\begin{center}
%\subfigure[Initial state  $I_{1x}$]{\includegraphics[scale=.1]{I1x.jpg}}
\subfigure[Constant $m_z$ ]{\includegraphics[scale=.15]{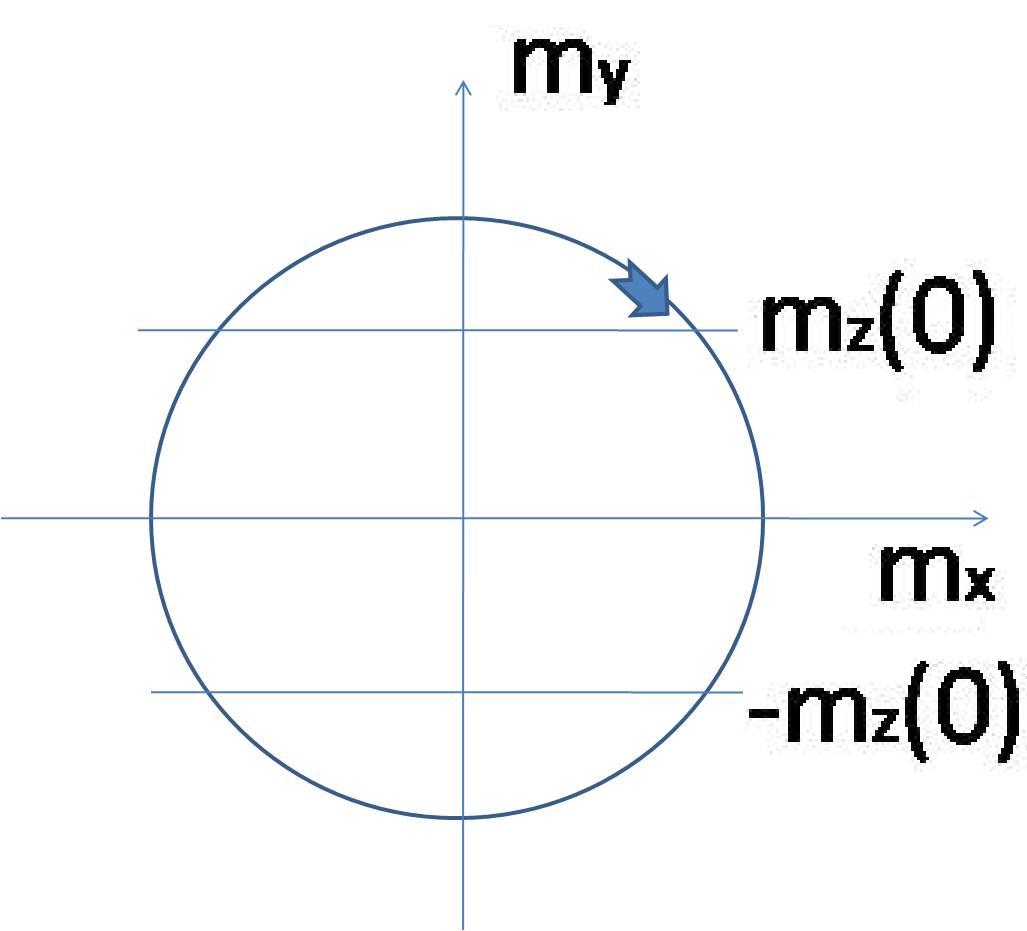} \label{fig:m1}}
%\end{center}
%\caption{ Trajectory}
%   \label{fig:m1}
%\end{figure}
\subfigure[Constant $m_y$]{\includegraphics[scale=.15]{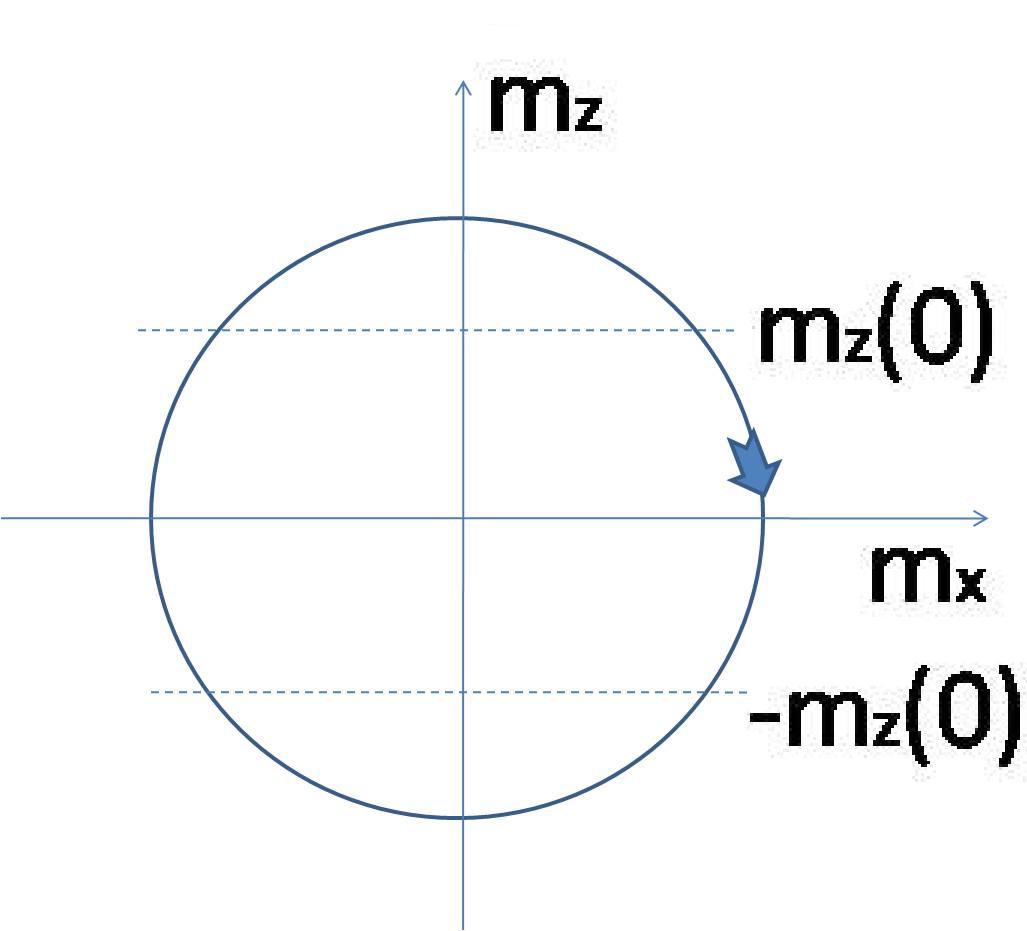}\label{fig:m2}}\\
\subfigure[Constant $m_z$]{\includegraphics[scale=.15]{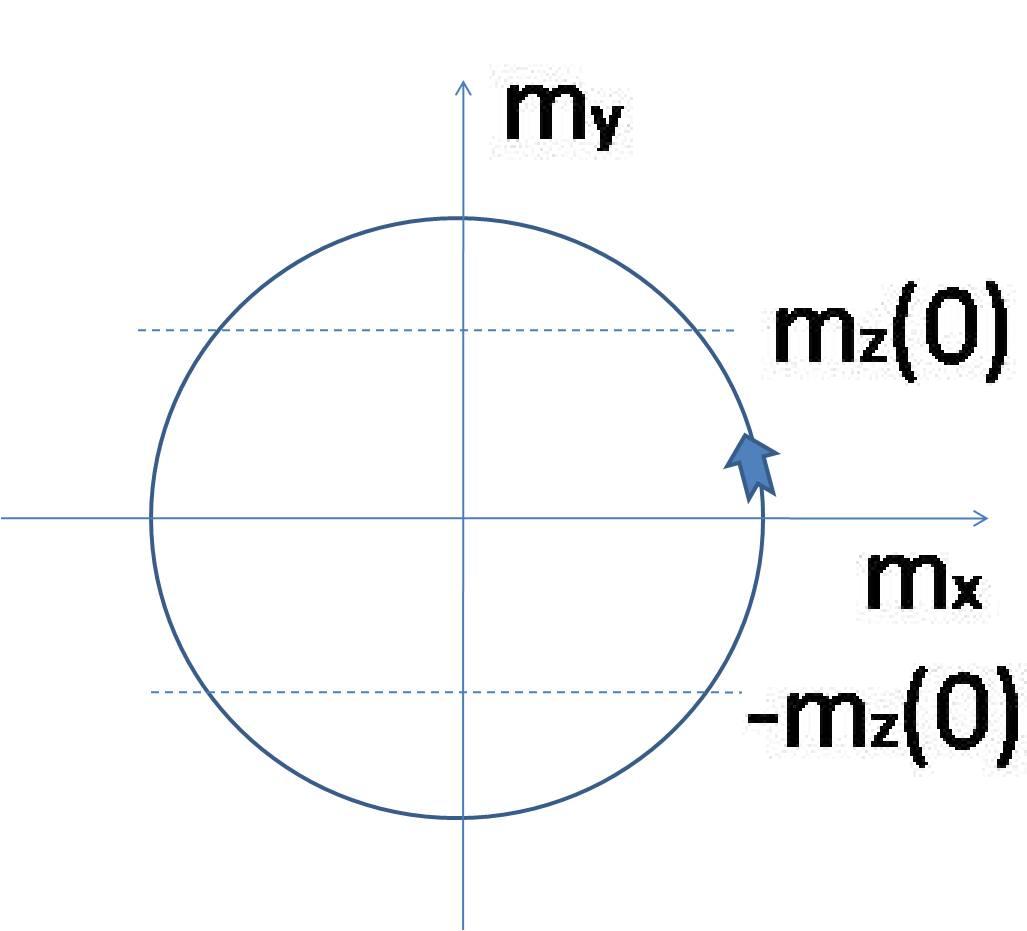}\label{fig:m3}}
\subfigure[Constant $m_y$]{\includegraphics[scale=.15]{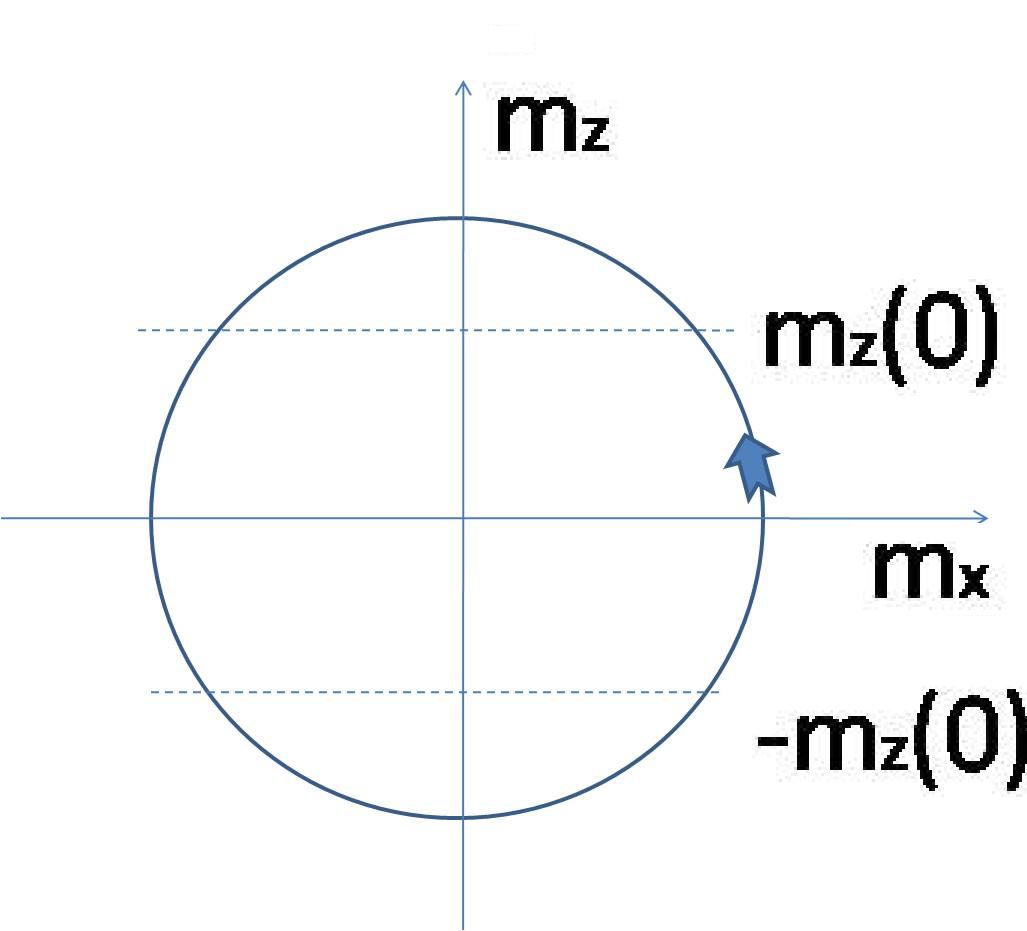}\label{fig:m4}}
\end{center}
\caption{ The evolution trajectory of $m_x,m_y,m_z$. }
%   \label{fig:m2}
\end{figure}

Assume that initially $m_z(0)>m_y(0)>0, m_x(0)>0$ (solutions under other initial conditions can be similarly worked out), then initially $$u=-sgn(m_z(0))=-1, v=0$$
From Eq.~\ref{eq:M} we get
\begin{eqnarray}
\label{eq:M1}
\aligned
\dot{m_x}&=m_y,\\
\dot{m_y}&=-m_x,   \\
\dot{m_z}&=0.\\
\endaligned
\end{eqnarray}
It evolves as in Fig.\ref{fig:m1}.

\n Case $\mathbf{I}$: If $m_x^2(0)+m_y^2(0)<m_z^2(0)$, then the controls are constant, $u=-sgn(m_z(0))=-1, v=0$ through out. \\
Case $\mathbf{II}$:If $m_x^2(0)+m_y^2(0)>m_z^2(0)$,
%first $u=-sgn(m_z(0))=-1, v=0$,
%\begin{eqnarray}
%%\label{eq:M}
%\aligned
%\dot{m_x}&=m_y,\\
%\dot{m_y}&=-m_x,   \\
%\dot{m_z}&=0.\\
%\endaligned
%\end{eqnarray}
%assume $m_x(0)>0$,
then after evolving for some time, $|m_y|$ will exceed $|m_z|$, so we need to switch the controls at time point $t1$, where $m_y(t_1)=-m_z(0)$, to $v=-sgn(m_y)=1, u=0$, then from Eq.~\ref{eq:M} we get
\begin{eqnarray}
%\label{eq:M}
\aligned
\dot{m_x}&=m_z,\\
\dot{m_y}&=0,   \\
\dot{m_z}&=-m_x.\\
\endaligned
\end{eqnarray}

which evolves as in fig.\ref{fig:m2} until $m_z(t_2)=-m_z(0)$ at some point $t_2$, where we switch the controls to $u=-sgn(m_z)=1, v=0$ and the dynamics changes to
\begin{eqnarray}
%\label{eq:M}
\aligned
\dot{m_x}&=-m_y,\\
\dot{m_y}&=m_x,   \\
\dot{m_z}&=0.\\
\endaligned
\end{eqnarray}
%\begin{figure}[h]
%\begin{center}
%\includegraphics[scale=.5]{m3.jpg}
%\end{center}
%\caption{ Trajectory}
%   \label{fig:m3}
%\end{figure}
which evolves as in fig.\ref{fig:m3} until $m_y(t_3)=m_z(0)$ where we switch the controls to $v=-sgn(m_y)=-1, u=0$ and the dynamics becomes
\begin{eqnarray}
%\label{eq:M}
\aligned
\dot{m_x}&=-m_z,\\
\dot{m_y}&=0,   \\
\dot{m_z}&=m_x.\\
\endaligned
\end{eqnarray}
%\begin{figure}[h]
%\begin{center}
%\includegraphics[scale=.5]{m4.jpg}
%\end{center}
%\caption{ Trajectory }
%   \label{fig:m4}
%\end{figure}
which will evolves till $m_z(t_4)=m_z(0)$ at some time point $t_4$, where we switch back to Eq.(\ref{eq:M1}) and the process repeats.Other initial conditions of $m_x, m_y,m_z$ give similar periodical controls.
So the optimal sequences in this case have the following pattern:

\begin{displaymath}
\begin{array}{cccccccc}
%\aligned
&1 \qquad&2 \qquad&3 \qquad&4 \qquad&5 \qquad &\ldots\qquad &n \qquad \\
&t_1 &\delta t &\delta t &\delta t  &\delta t &\ldots &t_2\\
u,v &1,0 &0,-1 &-1,0 &0,1  &1,0 &\ldots &\\
u,v &1,0 &0,1 &-1,0 &0,-1  &1,0 &\ldots &\\
u,v &-1,0 &0,-1 &1,0 &0,1  &-1,0 &\ldots &\\
u,v &-1,0 &0,1 &1,0 &0,-1  &-1,0 &\ldots &
%\endaligned
\end{array}
\end{displaymath}
and also the sequences with $u,v$ switched. Here $t_1,t_2\leq \delta t< \pi, $ are the corresponding evolution periods of each step, the evolution periods for the intermediate steps are equal.

Case $\mathbf{III}$: If $m_x^2(0)+m_y^2(0)=m_z^2(0)$, the process starts similarly, first $u=-sgn(m_z(0))=-1, v=0$,
\begin{eqnarray}
%\label{eq:M}
\aligned
\dot{m_x}&=m_y,\\
\dot{m_y}&=-m_x,   \\
\dot{m_z}&=0.\\
\endaligned
\end{eqnarray}
%assume $m_x(0)>0$,
It evolves till $m_y(t_1)=-m_z(0)$ at some time point $t_1$. But since $m_x^2(0)+m_y^2(0)=m_z^2(0)$, so at time $t_1$, $m_x(t_1)=0$. From Eq.(\ref{eq:M}), at time $t_1$
\begin{eqnarray}
%\label{eq:M}
\aligned
\dot{m_x}&=-um_y+vm_z,\\
\dot{m_y}&=um_x=0,   \\
\dot{m_z}&=-vm_x=0.\\
\endaligned
\end{eqnarray}
This is a singular point and we can choose $u,v$ such that $\dot{m_x}=-um_y+vm_z=0$, which can be achieved by rapidly changing between $u=-1,v=0$ and $u=0,v=1$(equivalent to evolve along $\Omega_y-\Omega_z$), and the dynamics can stay at this point for arbitrary long time. After that, one can either continue with $u=-1,v=0$ or switch to $u=0, v=1$. If it continues with $u=-1,v=0$, then it will evolves till $t_2$ such that $m_y(t_2)=m_z(0)$ and reach another singular point, where we can switching rapidly between $u=-1,v=0$ and $u=0,v=-1$, which is equivalent to evolve along $-(\Omega_y+\Omega_z)$. If it continues with $u=0,v=1$, then it follows the dynamics
  \begin{eqnarray}
%\label{eq:M}
\aligned
\dot{m_x}&=m_z,\\
\dot{m_y}&=0,   \\
\dot{m_z}&=-m_x.\\
\endaligned
\end{eqnarray}
till $t_2$ such that $m_z(t_2)=-m_z(0)=m_y(t_2)$, where it can switching rapidly between $u=1,v=0$ and $u=0,v=1$, which is equivalent to evolve along $\Omega_y+\Omega_z$. So in this case we have following possible sequences
  \begin{eqnarray}
%\label{eq:M}
\aligned
e^{-\delta t_1\Omega_{z}}&e^{\delta t_2(\Omega_y-\Omega_z)}e^{- \pi\Omega_{z}}e^{-\delta t_3(\Omega_y+ \Omega_z)}...e^{\delta t_n \Omega_{z/y}},\\
e^{-\delta t_1\Omega_{z}}&e^{\delta t_2(\Omega_y-\Omega_z)}e^{\pi\Omega_{y}}e^{\delta t_3(\Omega_y+ \Omega_z)}...e^{\delta t_n\Omega_{z/y}}\\
&\vdots
\endaligned
\end{eqnarray}

Here $\Omega_{z/y}$ means it can be either $\Omega_z$ or $\Omega_y$. The sequence is switching between regular points($\Omega_{z/y}$) and singular points($\Omega_z\pm\Omega_y$) and the regular points in the middle of sequence have to evolve for $\pi$ units of time each. For the optimal sequence, we can just consider the sequences with at most one singular point, as if it appears twice, for example if the optimal sequence contains $e^{\delta t_2(\Omega_y-\Omega_z)}e^{ \pi\Omega_{y}}e^{\delta t_3(\Omega_y+\Omega_z)}$, then we can replace it by
  \begin{eqnarray}
%\label{eq:M}
\aligned
 &e^{\delta t_2(\Omega_y-\Omega_z)}e^{ \pi\Omega_{y}}e^{\delta t_3(\Omega_y+\Omega_z)}e^{ -\pi\Omega_{y}}e^{\pi\Omega_{y}}\\
 =&e^{(\delta t_2+\delta t_3)(\Omega_y-\Omega_z)}e^{\pi\Omega_{y}},
\endaligned
\end{eqnarray}
 which has only one singular point with same time cost.
% thus it turns out that these sequences are actually not optimal for our case.

With these information, we can determine the optimal sequence to generate $e^{\alpha\Omega_x}, \alpha \in [0,\pi/2]$.
The element $\{e^{\alpha\Omega_x}\}$ has a one to one correspondence to the action on sphere of rotating the point $(0, \sin\alpha, \cos\alpha)$ to (0,0,1) while keeping the x-axis fixed. We now study the time optimal way to generate such actions.

Clearly the constant controls can't rotate the point $(0, \sin\alpha, \cos\alpha)$ to $(0,0,1)$ and using two rotations we can't rotate the point $(0, \sin\alpha, \cos\alpha)$ to $(0,0,1)$ while keeping the x-axis fixed. There are a few ways to generate $e^{\alpha\Omega_x}$ using three rotations, they are all essentially equivalent to $$e^{\alpha\Omega_z}e^{\beta (\Omega_z+\Omega_y)}e^{\gamma\Omega_z},$$
$$e^{\alpha\Omega_z}e^{\beta (\Omega_z+\Omega_y)}e^{\gamma\Omega_y},$$ or
$$e^{\alpha\Omega_z}e^{\beta\Omega_y}e^{\gamma\Omega_z}.$$
All these pulses turn out taking longer time than the four rotations presented below, so we will not present the detail of the calculation on these pulses.
\begin{figure}[h]
\begin{center}
\includegraphics[scale=0.3]{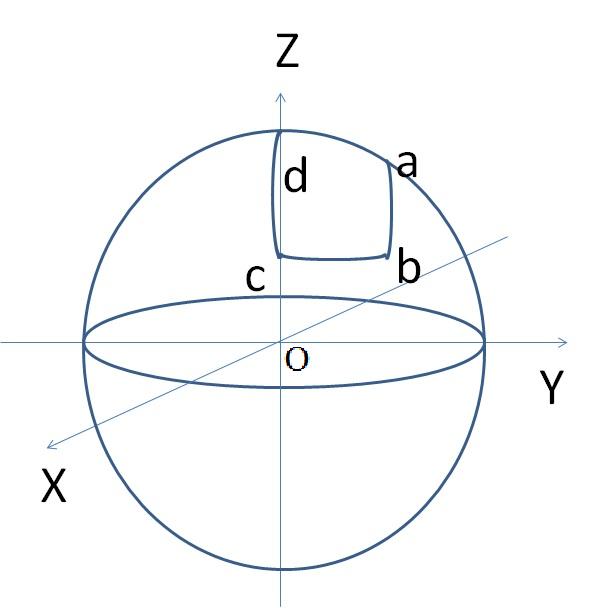}
\end{center}
\caption{ Rotations that moves $a=(0, \sin\alpha, \cos\alpha)$ to $d=(0,0,1)$ via the point $b=(\cos\alpha \sin t_1, \sin\alpha, \cos\alpha \cos t_1)$ and $c=(\sqrt{\cos^2\alpha \sin^2 t_1+\sin^2\alpha},0,\cos\alpha \cos t_1)$.}
   \label{fig:rotation1}
\end{figure}

We now give the strategy with four rotations that moves $(0, \sin\alpha, \cos\alpha)$ to $(0,0,1)$ and keeps the x-axis fixed. As shown in Fig.\ref{fig:rotation1}, it is first rotated around the y-axis for $t_1$ time, which rotates the point $a=(0, \sin\alpha, \cos\alpha)$ to $b=(\cos\alpha \sin t_1, \sin\alpha, \cos\alpha \cos t_1)$; then it is rotated around (-z)-axis for $\delta t$ time, which rotates point b to point $c=(\sqrt{\cos^2\alpha \sin^2 t_1+\sin^2\alpha},0,\cos\alpha \cos t_1)$ which is on the XZ-plane; after that it is rotated around the (-y)-axis for $\delta t$ time to the point $d=(0,0,1)$; since the x-axis is orthogonal to $\overrightarrow{oa}$ and rotations does not change the angles, so after these rotations, the x-axis has moved to somewhere which is orthogonal to $\overrightarrow{od}$, i.e., somewhere on the XY-plane, so we need to make another rotation around the z-axis for $t_2$ time, which moves the x-axis back to the original position. After combining these rotations, we get
\begin{equation}
e^{\alpha \Omega_x}=e^{t_2\Omega_z}e^{-\delta t \Omega_y}e^{-\delta t\Omega_z}e^{t_1\Omega_y}.
\end{equation}
From the second and third rotations, we get two equations on $\delta t$ and $t_1$
\begin{eqnarray}
\aligned
tan(\delta t)&=\frac{\sin\alpha}{\cos\alpha \sin t_1},\\
tan(\delta t)&=\frac{\sqrt{\cos^2\alpha \sin^2 t_1+\sin^2\alpha}}{\cos\alpha \cos t_1}.
\endaligned
\end{eqnarray}
which we can solve to get the value of $t_1$ and $\delta t$:
\begin{eqnarray}
\aligned
t_1&=arccos\frac{1}{\sin\frac{\alpha}{2}+\cos\frac{\alpha}{2}},\\
\delta t&=arccos(\cos\frac{\alpha}{2}-\sin\frac {\alpha}{2}).
\endaligned
\end{eqnarray}
To calculate the value of $t_2$, we need to figure out the trajectory of the x-axis: it is first rotated from $(1,0,0)$ to $(\cos t_1,0, -\sin t_1)$, then it is rotated to $(\cos t_1 \cos\delta t, -\cos t_1 \sin\delta t, -\sin t_1)$, then to $(\sqrt{\\cos^2t_1 \cos^2\delta t+\sin^2 t_1}, -\cos t_1 \sin \delta t, 0)$. The last rotation around z-axis should rotate $(\sqrt{\cos^2t_1 \cos^2\delta t+\sin^2 t_1}, -\cos t_1 \sin \delta t, 0)$ back to (1,0,0), i.e.,
$$\cos t_2=\sqrt{\cos^2t_1 \cos^2\delta t+\sin^2 t_1}=\frac{1}{\sin\frac{\alpha}{2}+\cos\frac{\alpha}{2}},$$
so $$t_2=t_1=arccos\frac{1}{\sin\frac{\alpha}{2}+\cos\frac{\alpha}{2}}.$$
So the total time to generate $e^{\alpha \Omega_x}$ is
\begin{equation}
%\label{eq:Time}
f(\alpha)=2[arccos\frac{1}{\sin\frac{\alpha}{2}+\cos\frac{\alpha}{2}}+arccos(\cos\frac{\alpha}{2}-\sin\frac {\alpha}{2})].
\end{equation}
Symmetrically when $\alpha\in [-\frac{\pi}{2},0]$, $$f(\alpha)=f(-\alpha).$$ So for $\alpha \in [-\frac{\pi}{2},\frac{\pi}{2}]$,
\begin{equation}
f(\alpha)=2[arccos\frac{1}{\sin\frac{|\alpha|}{2}+\cos\frac{|\alpha|}{2}}+arccos(\cos\frac{|\alpha|}{2}-\sin\frac {|\alpha|}{2})].
\end{equation}
This is actually the optimal sequences, adding more switches would not help\cite{Mittenhuber}. For completeness, we sketch the proof in \cite{Mittenhuber} showing a representative sequence $e^{t\Omega_y}e^{s\Omega_z}e^{-s\Omega_y}e^{-s\Omega_z}e^{t\Omega_y}$ can not be part of optimal sequence, thus optimal sequences can not have more than four rotations. First note that $$e^{s\Omega_z}e^{-s\Omega_y}e^{-s\Omega_z}=e^{-(\pi-s)\Omega_z}e^{s\Omega_y}e^{(\pi-s)\Omega_z},$$ as $3s>2\pi -s$ when $s>\frac{\pi}{2}$, thus only for  $s\leq \frac{\pi}{2}$, $e^{t\Omega_y}e^{s\Omega_z}e^{-s\Omega_y}e^{-s\Omega_z}e^{t\Omega_y}$ can be time optimal.

For $s=\frac{\pi}{2}$,
\begin{eqnarray}
\aligned
e^{t\Omega_y}e^{\frac{\pi}{2}\Omega_z}e^{-\frac{\pi}{2}\Omega_y}e^{-\frac{\pi}{2}\Omega_z}&=e^{t\Omega_y}e^{-\frac{\pi}{2}\Omega_x}\\
&=e^{t\Omega_y}e^{-\frac{\pi}{2}\Omega_y}e^{\frac{\pi}{2}\Omega_z}e^{\frac{\pi}{2}\Omega_y}\\
&=e^{(t-\frac{\pi}{2})\Omega_y}e^{\frac{\pi}{2}\Omega_z}e^{\frac{\pi}{2}\Omega_y},\\
\endaligned
\end{eqnarray}
as $\frac{\pi}{2}-t+\frac{\pi}{2}+\frac{\pi}{2}<t+\frac{3\pi}{2}$, thus the sequences with $s=\frac{\pi}{2}$ can not be optimal.

For $s\in (0,\frac{\pi}{2})$, it can be shown that for $t$ sufficiently small, there exists $r_1(t),r_2(t),r_3(t)$ such that $$e^{t\Omega_y}e^{s\Omega_z}e^{-s\Omega_y}e^{-s\Omega_z}e^{t\Omega_y}=e^{r_1\Omega_y}e^{r_2\Omega_z}e^{-r_2\Omega_y}e^{-r_3\Omega_z},$$
and $r_1+2r_2+r_3<2t+3s$. Thus $e^{t\Omega_y}e^{s\Omega_z}e^{-s\Omega_y}e^{-s\Omega_z}e^{t\Omega_y}$ can not be optimal or part of optimal sequences.
%As for any $s\in (0,\frac{\pi}{2})$, let $$\gamma_s(t)=e^{t\Omega_y}e^{s\Omega_z}e^{-s\Omega_y}e^{-s\Omega_z}e^{t\Omega_y}.$$ Consider the map $F:\R^3\rightarrow SO(3)$ given by $F(r_1,r_2,r_3)=e^{r_1\Omega_y}e^{r_2\Omega_z}e^{-r_2\Omega_y}e^{-r_3\Omega_z},$
% then $\gamma_s(0)=F(0,s,s)$.
%\section{Simulation of NMR spectra}
\begin{figure}[ht]
\label{fig:sim}
\begin{center}
\subfigure[Initial state  $I_{1x}$]{\includegraphics[scale=.5]{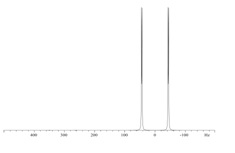}}
\subfigure[Final state $I_{1z}I_{2z}I_{3x}$]{\includegraphics[scale=.5]{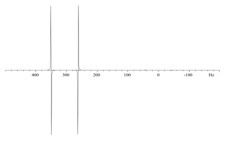}}\\
\subfigure[Initial state  $I_{1y}$]{\includegraphics[scale=.5]{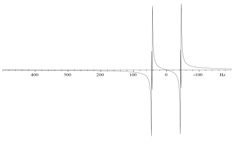}}
\subfigure[Final state $-I_{1z}I_{2z}I_{3y}$]{\includegraphics[scale=.5]{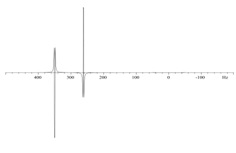}}\\
\subfigure[Initial state  $I_{3x}$]{\includegraphics[scale=.5]{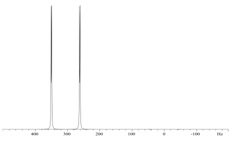}}
\subfigure[Final state $I_{1x}I_{2z}I_{3z}$]{\includegraphics[scale=.5]{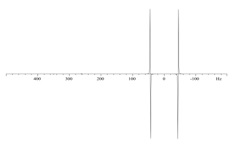}}\\
\subfigure[Initial state  $I_{3y}$]{\includegraphics[scale=.5]{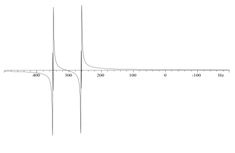}}
\subfigure[Final state $-I_{1y}I_{2z}I_{3z}$]{\includegraphics[scale=.5]{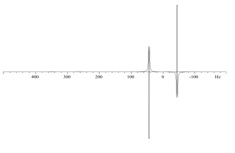}}\\
\subfigure[Initial state  $I_{1z}$]{\includegraphics[scale=0.5]{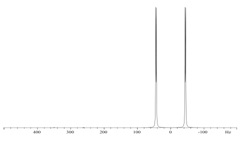}}
\subfigure[Final state $-I_{3z}$]{\includegraphics[scale=.5]{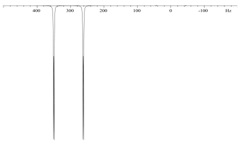}}\\
\subfigure[Initial state  $I_{3z}$]{\includegraphics[scale=.5]{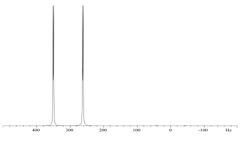}}
\subfigure[Final state $-I_{1z}$]{\includegraphics[scale=.5]{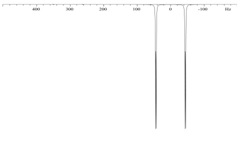}}
%
%
%
%
%\begin{subfigure} [h]{0.1pt}
%\centering
%\includegraphics[scale=.2]{I1x.jpg}
%\caption{Initial state $I_{1x}$}
%\end{subfigure}
%~
%\begin{subfigure} [h]{0.1pt}
%\centering
%\includegraphics[scale=.2]{I1zI2zI3x.jpg}
%\caption{Final state $4I_{1z}I_{2z}I_{3x}$}
%\end{subfigure}
\end{center}
\caption{Simulation of NMR spectra of different initial and final states, where left column are spectra of different initial states and right column are spectra of corresponding output states.}
\end{figure}

\end{document}